\documentclass[abbrv,aps,prb,twocolumn,showpacs,preprintnumbers,amsmath,amssymb,
superscriptaddress,nofootinbib]{revtex4}
\usepackage{graphicx}
\usepackage{dcolumn}
\usepackage{color}

\begin{document}

\title{Electrodynamics of BaFe$_2$As$_2$ from infrared measurements under pressure}
\author{L.~Baldassarre}
\author{A.~Perucchi}
\affiliation{Sincrotrone Trieste, Area Science Park, 34012 Trieste, Italy}
\author{P.~Postorino}
\affiliation{CNR-IOM and
Dipartimento di Fisica, Universit\`a di Roma "Sapienza", Piazzale
Aldo Moro 2, 00185 Rome, Italy}
\author{S.~Lupi}
\affiliation{Sincrotrone Trieste, Area Science Park, 34012 Trieste, Italy}
\affiliation{CNR-IOM and
Dipartimento di Fisica, Universit\`a di Roma "Sapienza", Piazzale
Aldo Moro 2, 00185 Rome, Italy}
\author{C.~Marini}
\affiliation{CNR-IOM and
Dipartimento di Fisica, Universit\`a di Roma "Sapienza", Piazzale
Aldo Moro 2, 00185 Rome, Italy}
\affiliation{European Synchrotron Radiation Facility, BP 220, 38043 Grenoble 
Cedex, France}
\author{L.~Malavasi}
\affiliation{Dipartimento di Chimica Fisica M. Rolla, INSTM (UdR Pavia) 
and IENI-CNR, Universit\`a di Pavia, Viale Taramelli 16,
27100 Pavia, Italy}
\author{J.~Jiang} 
\author{J.D.~Weiss} 
\author{E.E.~Hellstrom}
\affiliation{Applied Superconductivity Center,
National High Magnetic Field Laboratory, Florida State University,
2031 East Paul Dirac Drive, Tallahassee, FL 32310, USA}
\author{I.~Pallecchi}
\affiliation{CNR-SPIN, Corso Perrone 24, 16152 Genova, Italy}
\author{P.~Dore}
\affiliation{CNR-SPIN and Dipartimento di
Fisica, Universit\'{a} di Roma "Sapienza", P.le A.Moro 2, 00185
Rome, Italy}

\begin{abstract}
We report on an infrared study on the undoped compound BaFe$_2$As$_2$ as a function 
of both pressure (up to about 10 GPa) at three temperatures (300, 160, and 110 K). 
The evolution with pressure and temperature of 
the optical conductivity shows that, by increasing pressure, the mid-infrared 
absorptions associated with magnetic order are lowered while the Drude term increases, indicating the
evolution towards a conventional metallic state. We evaluate the 
spectral weight dependence on pressure comparing it to that 
previously found upon doping. The whole optical results indicate
that lattice modifications can not be recognized as the only parameter determining 
the low-energy electrodynamics in these compounds.

\end{abstract}


\maketitle
\section{Introduction}
The discovery of superconductivity in pnictides\cite{Kami} has 
triggered tremendous interest in a large scientific community.
Despite the large experimental and theoretical efforts\cite{review1,review2}, 
many questions are left unanswered, in particular concerning the subtle 
interplay among electronic, magnetic, and structural degrees of freedom. 

The common characteristic of these systems is that the undoped parent compounds 
have a metallic behavior, and undergo a transition from a high temperature 
paramagnetic phase to an antiferromagnetic spin-density-wave (SDW) state at 
the temperature $T_{SDW}$ ($\sim$ 150 K), accompanied by a tetragonal to 
orthorhombic (or monoclinic) structural transition. 
Upon entering the SDW state, the Fermi surface of these 
multi-band systems is only partially gapped, therefore the metallic state 
persists in spite of the reduction of the carrier density.  
Superconductivity is achieved, below $T_c$, 
either by electron (or hole) doping or, in the undoped parent compounds, 
by applying external pressure \cite{Chu2009}. Furthermore, in the case of 
superconducting systems, the applied pressure can remarkably enhance 
$T_c$ \cite{Ahilan2009}. 
It is now known that structure and properties of chemically doped samples are 
strongly linked\cite{rotter,zhao}, for example a decrease of the Fe-Fe distance 
(that occurs upon chemical doping) has been found to strongly enhance $T_c$. 
A similar lattice modification was found also in the case of applied 
pressure\cite{Kim2009} and the idea has been put forward that the structural 
changes play a major role in determining the electrodynamic properties. 
It is thus clear that a complex interplay among different degrees of freedom 
is at work in these systems\cite{Yin}, which makes challenging the understanding of 
their intriguing physical properties. 

The complex phase diagram of pnictides has been carefully addressed in particular 
in systems of the so-called 122 family $A$Fe$_2$As$_2$ ($A$ = Ba, Sr, Ca, Eu): 
superconductivity is here obtained either by chemical substitutions in the $A$ 
plane (hole doping) or in the Fe-As layer (electron doping), and large effects 
have been reported as a function of physical pressure\cite{Kim2009,Ahilan2009,Duncan2010}. For the undoped compounds, 
in particular, it has been shown that pressure 
progressively lowers $T_{SDW}$, eventually suppressing the 
structural/magnetic transition, and can originate the insurgence 
of the superconducting phase at the cost of the SDW state. The same effect can be achieved by means of a chemical pressure by replacing smaller P- atoms for the larger As-atoms, with no effect of charge doping.\cite{cao}
Most authors agree on the fact that chemical and physical pressure act similarly on these compounds, however it was recently pointed out that in Co-doped compounds the evolution of the Fe-As distance is different\cite{zinth} with physical and chemical pressure (by replacing As with P). As this parameter is considered to be crucial in determining the properties of pnictides, a deeper insight in the equivalence doping-pressure is thus required.

Several infrared studies have been performed on Iron-based superconductors 
to probe both the normal and the superconducting state\cite{SUST,Dres_gap, perucchi, Wu2010}. 
In 122 systems, in particular, studies have been performed
in order to asses the effect of the magnetic transition at $T_{SDW}$ on the 
electronic structure. A general consensus has been achieved on the existence 
in the $ab$-plane optical response of at least two electronic subsystems, that 
can be described by two Drude contributions in the frequency dependent 
conductivity. 
In these two contributions, characterized by different scattering rates, 
the transition to the low-temperature SDW state can origin the opening of 
energy gaps with transfer of optical spectral weight from low to high frequencies 
\cite{Wu2010, Chen2010}. 

Recently particular attention has been devoted to the study of the degree of 
electronic correlations\cite{Basov_manifesto, quazilbash, Schafgans}, that is the ratio 
between the experimental kinetic energy ($K_{exp}$) and that obtained through band
calculations ($K_{theory}$) often performed in the Local Density Approximation.
A value $K_{exp}/K_{theory}$ close to 1 indicates a \emph{conventional} metallic 
state (\emph{i.e.} with no tendency towards charge localization due to correlations) while 
the opposite limit  $K_{exp}$/$K_{theory}$ $\rightarrow0$ 
identifies a system with fully localized electrons.
Although this interpretation of $K_{exp}/K_{theory}$ parameter has to be taken with 
some caution\cite{ arXiv_0910_4117, Schafgans}, 
especially in multi band systems, it has been shown that \cite{Basov_manifesto} 
there is a similar tendency towards localization\cite{Basov_manifesto, Schafgans} 
in different families of high-$T_c$ and that this parameter might be
a key ingredient in high-$T_c$ superconductivity\cite{Basov_manifesto, Schafgans}. 
In the above sketched scenario, infrared measurements are of basic 
interest since the integral of the Drude contribution to the frequency dependent 
conductivity (i.e. the Drude spectral weight) simply mimics the kinetic energy 
of the free carries. 
Moreover, an infrared study performed as a function of temperature and pressure 
on undoped pnictides, avoiding disorder or local distortions induced by doping, 
yields information on the pressure-driven changes in the frequency dependent 
conductivity, i.e. in the charge dynamics. 

In the present work, we report on an infrared investigation on the undoped 
compound BaFe$_2$As$_2$ as a function of pressure at three temperatures, 110K, 160K and 300K, \emph{i.e.} above and below T$_{SDW}$ at ambient condition. 
By evaluating the evolution of the infrared spectral weight with pressure, and 
comparing it to that found upon doping, we find that lattice 
modifications can not be recognized as the only parameter determining the 
low-energy electrodynamics in these compounds.

\section{Experimental details}
Infrared reflectance measurements were performed on high quality BaFe$_2$As$_2$ 
single crystal ($T_{SDW}$ = 140 K at ambient pressure). The crystals were prepared by the self-flux method\cite{JJ}.  
Ba pieces, Fe powder and As particles were mixed together according to the ratio of Ba:Fe:As=1:4:4. After mixing, the powder was wrapped with Nb foil and placed in a stainless steel ampoule. The ampoule was evacuated, welded shut, and compressed with a cold isostatic press at 275 MPa to press the powder into a pellet.  The heat treatment was done in a hot isostatic press at 193 MPa. The samples were hold for 9 hours at 1120°C, and ramped down to 910 °C at 2 °C/h and then cooled to room temperature in a hour. The pressure was released at the end of the heat treatment. The single crystals were extracted by breaking the reacted pellet.
 Data were collected on increasing pressure $P$ up to a value of 
10 GPa, at different temperatures $T$ between room-$T$ and 110 K.

Pressure-dependent reflectivity measurements were performed with a Diamond 
Anvil Cell (DAC). A 50 $\mu$m thick CuBe gasket was placed in a screw-driven 
opposing-plate DAC, equipped with type IIa diamonds. A small piece of the 
sample has been cut and placed over a pre-sintered CsI pellet in the hole 
drilled in the gasket. The good quality of surface obtained by breaking the 
sample allowed to have a clean sample-diamond (s-d) interface. 
The DAC was mounted inside a N$_2$-flow microscope cryostat from Oxford, 
allowing us to cool the DAC down to $\approx$ 100 K. 
In order to determine both pressure and temperature of the sample, a ruby 
chip was placed inside the sample chamber while a second one was placed on 
the external face of the diamond. The pressure value was measured in situ 
by the standard ruby fluorescence \cite{Mao} technique. 
Since the shift of the two fluorescence lines depends strongly on both 
$T$ and $P$, measurements were performed separately on the 
two ruby chips: on the external one to determine $T$ and on the 
internal for $P$. We remark that, by measuring the $T$-induced shift 
of the ruby line, we determine precisely the temperature of the sample 
which is in tight contact with the highly conducting diamond anvil.

The spacing, the linewidth, and the deep hole between the two ruby fluorescence lines are rather delicate markers of the hydrostaticity of the sample environment. Those have been mpnitored up to the highest measured pressure, ensuring us to be in reasonably good hydrostatic conditions. 

The incident and reflected light were focused and collected with an optical 
microscope mounting Cassegrain objectives and equipped with 
a Mercury- Cadmium- Telluride (MCT) detector and a bolometer. The microscope was coupled to the IFs66/v 
Bruker Michelson interferometer. This measuring configuration allowed us to 
explore the 300-12000 cm$^{-1}$ spectral range\cite{eprhg}. Due to the small 
size of the sample, the high brilliance of Synchrotron Radiation at SISSI 
beamline at ELETTRA storage ring \cite{sissi} was exploited, with a great 
advantage especially  for the measurements in the spectral region of the 
far-infrared.  The measurement  procedure was the same as described 
in Refs. \onlinecite{v3o5,lavagnini}.

\section{Results and discussion}
\begin{figure}
\includegraphics[width=0.9\columnwidth,angle=0,clip]{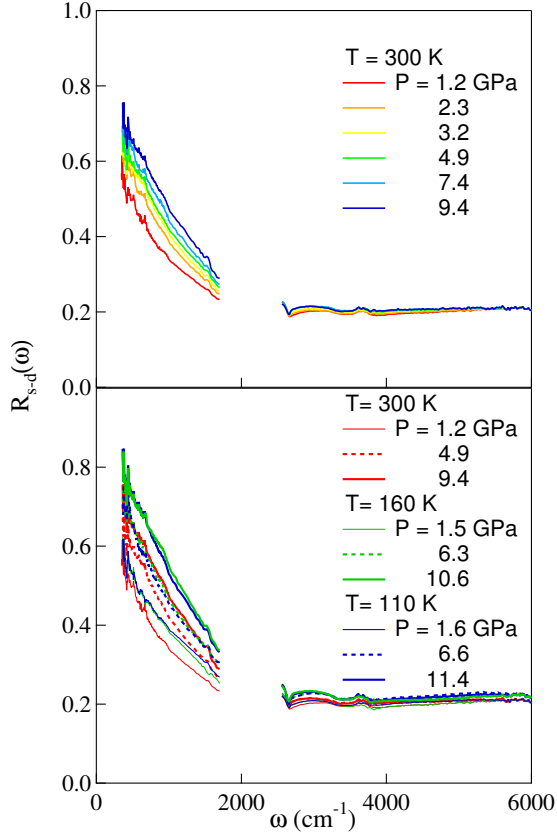}
\caption{\label{R}(Color online) (Upper Panel): Room temperature reflectivity at 
the sample-diamond interface shown in the 0-6000 cm$^{-1}$ frequency range 
for different pressures (indicated in figure). 
(Lower Panel): Reflectivity at the sample-diamond interface along three 
pressure-temperature cycles. Upon cooling down the pressure value increases 
significantly and sets back to the starting value once reached 
room temperature again.} 
\end{figure}

The reflectivity at the sample-diamond interface ($R_{s-d}$) is shown in Fig.\ref{R} 
in the 0-6000 cm$^{-1}$ frequency range for various pressures and 
temperatures. Data are not shown in the 1600-2700 cm$^{-1}$ due to 
the strong phonon absorption of the diamonds\cite{dore}. All curves merge at about 
6000 cm$^{-1}$, showing no pressure or temperature 
dependence above such frequency. At room temperature, 
all curves increase for decreasing frequency suggesting a metallic-like 
behavior and, by applying pressure, the reflectivity is enhanced in the 
whole energy range pointing to an increased metallic behavior. 
As $T<T_{SDW}$ ($P$=1.6 GPa, $T$=110K) $R_{s-d}$ increases its value in the mid-IR but shows only a small increase at low-frequency 
almost crossing the reflectivity curve recorded for $P$=1.2 GPa and $T$=300K. 

\begin{figure}
\includegraphics[width=0.95\columnwidth,angle=0,clip]{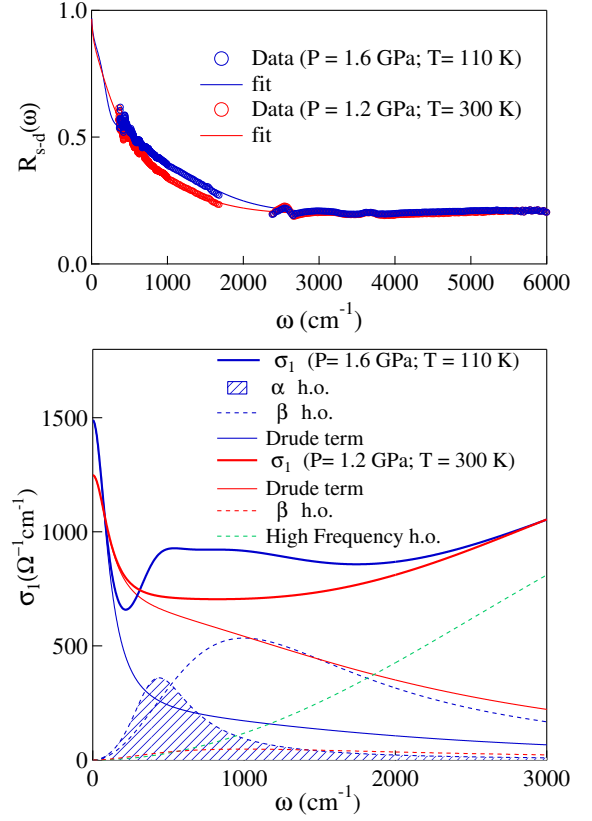}
\caption{\label{fit}(Color online) (Upper Panel): Reflectivity at the sample diamond interface 
compared with the best Drude-Lorenz fit. (Lower Panel): Optical conductivity 
$\sigma_1(\omega)$ obtained from the fitting procedure for $T$=110 K and 300 K.
The fit components are also reported: the sum of the two Drude terms and the 
harmonic oscillators (h.o.). The high frequency oscillator (that mimics the 
contributions above the measured frequencies) is kept constant for all the
measured curves.}
\end{figure}

\begin{figure*}
\includegraphics[width=1.4\columnwidth,angle=0,clip]{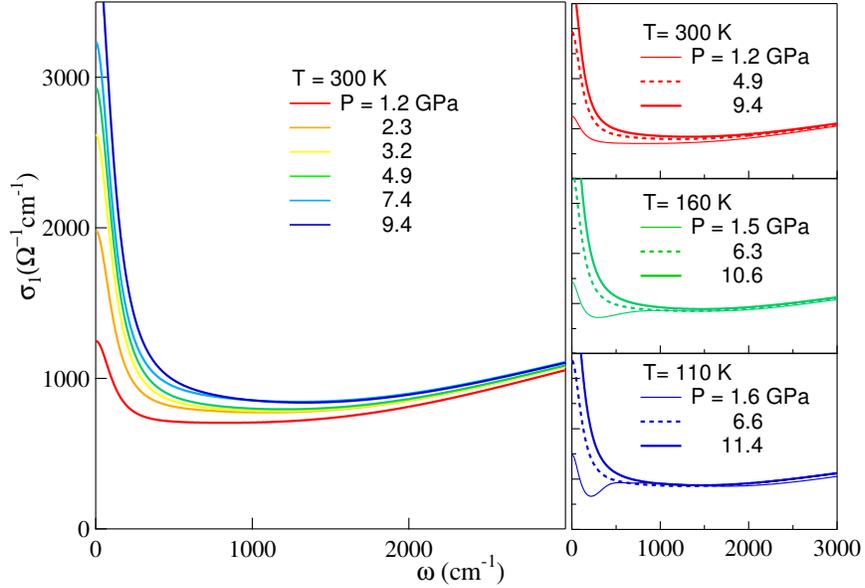}
\caption{\label{sigma}(Color online) (Left): Optical conductivity $\sigma_1(\omega)$ at various 
pressures at room temperature. As pressure is increased the conductivity increases
in the whole shown range. (Right): $\sigma_1(\omega)$ at 300, 160 and 110 K. 
At the lowest measured pressure (1.6 GPa), as the SDW state sets in, 
$\sigma_1(\omega)$ shows a depletion and a peak respectively below and slightly 
above 500 cm$^{-1}$ due to the partial gapping of the Fermi surface. At lower 
frequency it is possible to see the onset of a narrow Drude.} 
\end{figure*}

\begin{figure}
\includegraphics[width=0.9\columnwidth,angle=0,clip]{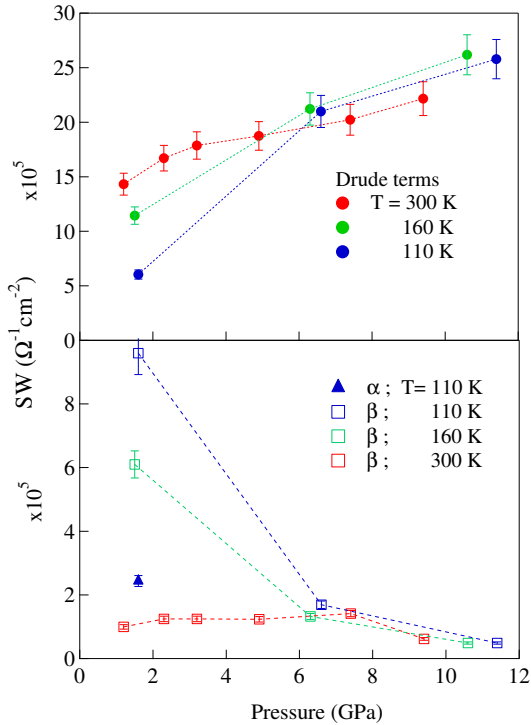}
\caption{\label{SW}(Color online) Pressure dependent changes in the spectral weight (SW) 
at 110 K (blue), 160 K (green) and 300 K (red) for the Drude terms (upper panel) 
and the mid-IR harmonic oscillator (lower panels). The lines are 
only a guide for the eye.} 
\end{figure}

To obtain the optical conductivity $\sigma_1(\omega)$ (real part of the complex
conductivity) we have fitted the reflectivity with a Drude-Lorenz model, taking 
into account the sample-diamond interface.\cite{plaskett} 

This has been done by first 
modeling the refractive index $\tilde{n}=n+ik$ as follows:\cite{Burns}
\begin{equation}
\epsilon_1= n^2-k^2=\epsilon_\infty+\sum_i\frac{S_i^2\omega_i^2-\omega^2}{(\omega_i^2-\omega^2)^2 + \Gamma_i^2\omega^2}
\end{equation}
\begin{equation}
\epsilon_2= 2nk=\sum_i \frac{S_i^2\Gamma_i\omega}{(\omega_i^2-\omega^2)^2 + \Gamma_i^2\omega^2}
\end{equation}
where S$_i$, $\Gamma_i$ and $\omega_i$ are respectively the oscillator strength, its width and central frequency. The Drude contribution is obtained once $\omega_i=0$. 
The reflectivity is thus modeled by using:
\begin{equation}
R_{s-d}=|\frac{\tilde{n}-\tilde{n}_D}{\tilde{n}+\tilde{n}_D}|^2
\end{equation}
where $\tilde{n}_D$ is the refractive index of diamond from Ref.\onlinecite{dore}.

We considered two 
Drude terms and three harmonic oscillators (h.o.), as proposed by S.J. Moon 
and coworkers\cite{mazin} and successfully used by other 
authors\cite{Lucarelli, Hu2008}. A high frequency h.o. 
has been used to mimic the absorptions that take place above the highest 
measured frequency and has been kept constant for all curves. Two Drude 
terms - one narrow and one broad - are necessary to fit our data: noteworthy 
this decomposition of the spectra holds for all the compounds of the 122 
family\cite{Dressel,Lucarelli,mazin}. 
The lack of low-frequency data prevents in our case an unambiguous determination 
of the spectral weight (SW) of each Drude component, while the SW of their sum 
results to be well determined. 

Two oscillators (that we call $\alpha$ and $\beta$, similarly to the notation in 
Ref.\onlinecite{mazin}) are needed to achieve satisfactory fittings in the mid-IR 
region below the SDW transition, while only one ($\beta$) is enough for $T>T_{SDW}$.  
In Fig.\ref{fit} we show the fitting curves together with the experimental data 
for 300 and 110 K and 1.2 and 1.6 GPa, \emph{i.e.} in the normal and in the SDW 
state respectively. 
For both the states, the resulting optical conductivity and the corresponding 
components are shown in the lower panel of Fig.\ref{fit}: 
as the SDW state sets in, there is a depletion of the conductivity below 
500 cm$^{-1}$ and the formation of strong mid-IR absorption bands above 
this frequency. 
At much lower frequencies the onset of the narrow Drude is visible. The SDW 
transition therefore results in a spectral weight transfer from the 
Drude terms to the $\alpha$ and $\beta$ oscillators.

\begin{figure}
\includegraphics[width=0.95\columnwidth,angle=0,clip]{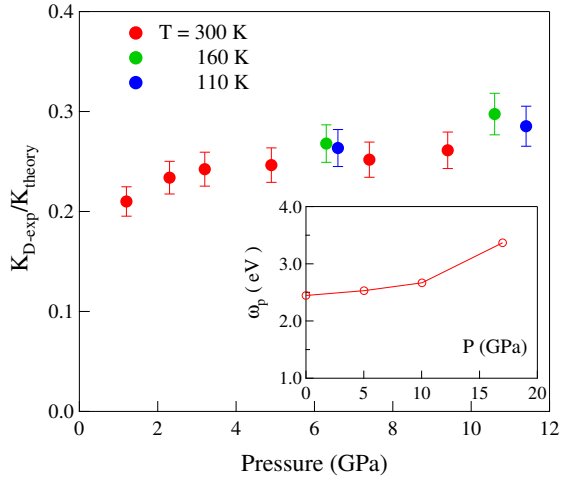}
\caption{\label{K}(Color online) $K_{D-exp}/K_{theory}$ plotted for increasing pressure in the 
paramagnetic phase. Plasma frequency values obtained by DFT 
calculations\cite{Profeta} are reported in the inset.} 
\end{figure}

The optical conductivity $\sigma_1(\omega)$ is plotted in Fig.\ref{sigma} at 
selected temperatures and pressures. At room temperature the conductivity 
increases its absolute value with increasing pressure, especially towards 
low frequency, indicating that the system tends towards a more metallic state. 
This is clearly evident from the pressure evolution of the Drude SW, shown in 
the upper panel of Fig.\ref{SW}. On the other hand, as pressure is increased, 
the SW of the absorption band in the mid-IR at room temperature remains almost 
constant and then decreases above 9 GPa. 
The SW of the $\beta$ h.o. decreases with increasing $P$ also at 110 and 
160 K as expected due to the progressive suppression of the magnetic state with 
pressure ($dT_{SDW}/dP<0$), as reported by several authors 
\cite{Ahilan2009,Mani,Ishi}. 
The SW of the $\beta$ oscillator in the mid-IR at 160 K and 
110 K, for $P>$ 6 GPa, is comparable with the SW attained at room temperature, 
suggesting that above this pressure the system is not undergoing any appreciable electronic 
phase transition. 
Once the SDW transition is suppressed, the external parameter that mainly 
drives the optical response is pressure, with only a small effect due to
temperature. Note that, by applying pressure, the overall SW increases and 
it is not recovered in the measured spectral range.

An accurate determination of the optical conductivity $\sigma_1(\omega)$ allows to 
follow the evolution upon external parameters of both free carrier response 
(Drude terms) and inter-band excitations (harmonic oscillators centered at finite 
frequency). 
The outcome of the fitting procedure determines also the 
integral of the Drude contribution to $\sigma_1(\omega)$, i.e. the Drude spectral weight. 
In order to evaluate the degree of correlation previously introduced, we considered 
the ratio between the experimental kinetic energy ($K_{D-exp}$) directly provided by 
the Drude spectral weight (see Ref.\onlinecite{quazilbash} for the complete formula),
and the theoretical kinetic energy ($K_{theory}$) given\cite{quazilbash} by the square 
of the plasma frequency $\omega_p$, obtained as a function of pressure in the paramagnetic 
phase by means of Density Functional 
Theory (DFT) computations\cite{Profeta}. Despite the light increase of $\omega_p$ with 
pressures up to 10 GPa, the resulting $K_{D-exp}$/$K_{theory}$ ratio increases 
smoothly on increasing pressure indicating that the system is shifting towards a conventional metallic state.  
It is worth noticing that the value that we find at the lowest measured 
pressure ($\sim 0.2$) is in good agreement with what is found (0.25-0.29) at ambient 
conditions by other authors\cite{Basov_manifesto,Schafgans}. 
A comparison between pressure and doping dependence of the degree of correlation is 
impossible, since its value was only reported for BaFe$_{1.84}$Co$_{0.16}$As$_2$ 
\cite{Schafgans} and its evolution with doping is not reported in literature.

\begin{figure}
\includegraphics[width=0.95\columnwidth,angle=0,clip]{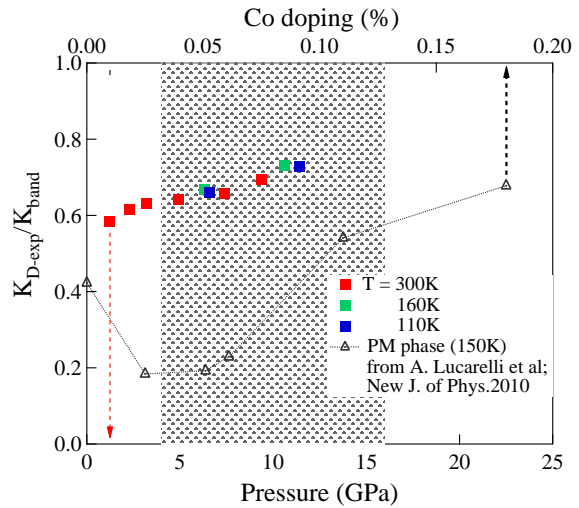}  
\caption{\label{Ksc} (Color online) $K_{D-exp}/K_{band}$ plotted for increasing pressure at 110, 
160 and 300 K  (all in the paramagnetic phase) compared with data from 
Ref.\cite{Lucarelli}. The gray shaded area identifies the regions (upon Co-doping 
and pressure) where superconductivity is attained.} 
\end{figure}

A different, fully experimental approach to evaluate the degree of correlation has 
been proposed by L. Degiorgi and coworkers\cite{Lucarelli,Degiorgi}: 
instead of comparing the experimental kinetic energy with a theoretical value, 
they discuss the ratio between the experimental kinetic energy given by the Drude 
spectral weight and the SW of $\sigma_1(\omega)$ including both Drude and mid-IR 
contributions ($K_{band}$). They find a degree of correlation $K_{D-exp}/K_{band}$ 
lower than 0.65 for the superconducting samples (Ba(Co$_x$Fe$_{1-x}$)$_2$As$_2$ with $x< 0.11$)
and higher ($\sim 0.7$) as the system becomes overdoped. 
In order to compare the effect of pressure and Co-doping on the degree of 
correlation we have also calculated the ratio $K_{D-exp}/K_{band}$. 
For this comparison we used the scaling relation 
proposed by S. Drotzinger et al.\cite{Drotz}: by means of transport and magnetic 
measurements in the superconducting region, they suggested 
$\Delta$P/$\Delta$x $\approx$ 1.275 GPa/$\%$Co.
In Fig.\ref{Ksc} we plot the $K_{D-exp}/K_{band}$ ratio for our data in the 
paramagnetic phase together with data (also in the paramagnetic phase) on Co-doped 
compounds (from Ref. \onlinecite{Lucarelli}).

We remark that we 
are not putting the accent on the differences  of the absolute value of this parameter 
(that could depend also on details of the high-frequency extrapolation  when performing 
KK analysis) but rather on its different evolution with doping and pressure. 
Supposing to start at the same value for undoped compound at ambient conditions, 
for low doping/pressure, chemical doping and pressure act differently: 
the former yields $K_{D-exp}/K_{band}$ values lower than those given 
by pressure, at odd with the fact that upon inserting Co in the FeAs planes does
not only modify the lattice parameters but also injects one carrier per dopant ion. 
It was shown that both pressure and doping suppress the magnetic/structural 
transitions but only certain dopants (\emph{i.e.} certain doping levels) allow 
superconducting pairing\cite{canfield_TM}. 
We know that upon doping BaFe$_2$As$_2$ with Co, for x$<$0.12, the SW of the mid-IR 
absorption bands is strongly increased, at odd with the SW of the Drude 
term\cite{Lucarelli}, suggesting that spin fluctuations
are enhanced while long range magnetic order is suppressed. 
This probably is one of the key factors in originating superconductivity once the 
structure is modified \emph{enough} due to the random insertion of Co ions. 
On the other hand, by applying pressure, the magnetic/structural
transition is suppressed but the SW of the mid-IR contributions is
reduced, suggesting that the magnetic fluctuations are lowered,
probably due to reduced anisotropy and weaker nesting
\cite{sacchetti}. The reduction of lattice parameters is enough for
quenching the SDW state (and, at lower temperatures, to originate
superconductivity), even though the system tends towards a fairly
conventional metallic state.
For x$>$ 0.12 the slope of $K_{D-exp}/K_{band}$ values for Co-doped
samples becomes similar to what we have measured under pressure,
suggesting that once the system is in the overdoped regime, a common
mechanism is driving the suppression of superconductivity.

In conclusion, we have studied the infrared response of BaFe$_2$As$_2$ under
pressure down to 110 K. We remark that pressure dependent IR measurements
can give the unique opportunity to go beyond the zero frequency conductivity,
providing informations on the low energy charge dynamics.
We observe no signature in the optical conductivity of SDW ordering in this temperature
range when P $\geq$ 3 GPa, in agreement with the results of previous transport
measurements.
It has been possible to follow the evolution with pressure of the spectral weight of
both Drude and mid-IR contributions, showing that the system tends towards a
conventional metallic state and that the absorptions associated with fluctuative
magnetic order are lowered by increasing pressure. On the basis of the
obtained results we could compare the effect of pressure and Co-doping on the
degree of correlation. The obtained results reported in Fig. 6 indicate that,
in the underdoped regime, pressure and Co-doping favor the onset of
superconductivity (destroy long-range magnetic order), by following
different mechanisms displaying opposite effects on the
degree of correlation parameter (see Fig.6). On the other hand, at high
pressure/doping the correlation degree monotonically increases in both
cases, thus suggesting that an increased itinerancy is the common
mechanism for $T_c$ reduction in the overdoped regime.
Our results indicate that the 
lattice distortions can not alone explain the evolution of the electrodynamics of 
underdoped compounds compared to that of the undoped sample under pressure since the routes 
taken to suppress the SDW and favour superconductivity are different in the 
two cases. A more realistic description of the interplay between different degrees of freedom 
is needed to understand the physics of pnictides.

\acknowledgements
We wish to thank M. Putti for providing the samples, G. Profeta and A. Continenza
for useful discussions and for providing $\omega_p$ values computed at a number of 
pressures prior to publication.
We acknowledge financial support from PRIN project No. 2008XWLWF9-002 and from CARIPLO 
Foundation (Project No. 2009-2540). The work at FSU was supported by NSF
DMR-1006584 and 0084173, and by the State of Florida.

  \end{document}